# Blockchain-Based Secure Online Voting Platform Ensuring Voter Anonymity, Integrity, and End-to-End Verifiability

Anthony Revilla
School Of Computing
Montclair State University
Fair Lawn, New Jersey
revillaa@montclair.edu

Yousef Tahboub
School Of Computing
Montclair State University
Montclair, New Jersey
tahbouby1@montclair.edu

Jaydon Lynch
School Of Computing
Montclair State University
Montclair, New Jersey
lynchj21@montclair.edu

Greg Floyd
School Of Computing
Montclair State University
Montclair, New Jersey
floydg1@montclair.edu

*Abstract*— **Casting a ballot from a phone or laptop sounds great but only if voters are sure their choice stays secret, and the results can't be changed in the dark. Our idea is to keep each encrypted vote on a small private blockchain run by whoever is organizing the vote (for example, a county election board, a company conducting a shareholder poll, or a student-government committee) plus observers. Every few minutes the system writes a short fingerprint (hashes in blockchain) of the newest votes on a public blockchain. The public hash works like a tamper-evident seal. Voters get a one-time digital ticket that hides their name plus a quick receipt showing their vote was accepted. This research explores that idea, positioning it as a middle-path approach between traditional voting methods and expensive, fully public blockchain models. While the research presented here is preliminary, the potential of building and testing such a system is significant, as it demonstrates a new way to balance accessibility, security, and transparency in democratic processes.**

## I. INTRODUCTION

Traditional in-person and mail voting both create headaches. Long lines, travel time, and postal delays discourage turnout, while lost envelopes or counting mistakes can cast doubt on close results. Moving the ballot box online could fix these issues, but most existing web-based systems depend on a single server. If that server is hacked or simply crashes on election day, ballots can be leaked, altered, or lost without anyone noticing. The fragility of centralized systems means that trust is easily broken, and in an election, trust is the most critical resource.

Several projects have tried putting votes on public blockchains so no single machine can cheat. While this makes tampering almost impossible, it also publishes every ballot transaction to the world and forces organizers to pay network fees that spike whenever the chain is busy. Privacy suffers, and costs become unpredictable. Voters should not have to decide between maintaining their anonymity and ensuring that their ballots are counted correctly.

We need a middle path, one that is convenient for voters, protects their anonymity, and stays affordable. Our idea stores encrypted ballots on a small, private blockchain run by the organizers plus a couple of neutral observers. Because the participants are known, the network reaches agreement quickly and charges no per-vote fees. After each batch of ballots is sealed, the system saves a tiny hash of that batch to a low-cost public blockchain. Anyone can later compare hashes to confirm nothing was added, removed, or changed.

To keep ballots private, each voter uses a one-time blind-signed ticket issued at check-in. The ticket proves the person is allowed to vote but never reveals who they are. Even the administrators who run the private ledger cannot link a stored ballot back to a name. This design ensures that the most important aspects of democracy—secrecy of the ballot, fairness of the count, and public verifiability—are upheld.

*Issues That Could Be Resolved*

• Voters often face long lines or travel requirements just to cast a ballot when they can just do it online easily and safely.

• Many current online systems lack strong security or privacy protection.

• Public blockchain methods can be too expensive or expose too much information. Establishing a decentralized model that includes trusted observers addresses both cost and privacy concerns.

## II. PROBLEM STATEMENT & MODEL

The reason this is an important issue is simple: if voting is hard, fewer will vote, and decisions that shape our communities, schools, and workplaces are left to a smaller, less representative group. Low turnout does not typically stem from people not caring about the issues at stake; instead, it often stems from practical barriers that block the way to the polls.

An hour-round-trip to a polling place, a queue that snakes round the block, or a single election day that interferes with work or caregiving responsibilities can turn an enthusiastic voter into a non-voter. Even mail voting, which is convenient, has its obstacles. A ballot can be lost in the mail stream, arrive too late for election officials to count, or be rejected for a minor signature discrepancy the voter never gets the chance to fix. All these friction points tally and are a form of silent disenfranchisement, skewing participation toward those with more relaxed work schedules, transportation, and fewer daily constraints. Over time, this erodes the credibility of democratic institutions because results fail to truly reflect the voice of the governed.

Online voting appears to offer an instant solution. Instead of juggling work schedules or hiring a babysitter, a voter can simply log on from home, a public library, or even a lunchtime smartphone. Travel time equals zero, weather no longer matters, and the window for voting can stay open for days rather than hours. For the physically or visually disabled, the features of accessibility in current devices, the screen readers, voice operation, and adjustable font sizes, present an independence that cannot be offered by traditional polling machines. On the administrative front, electronic ballots eliminate the expense and hassle of printing, shipping, and warehousing thousands of sheets of paper, slashing costs and streamlining logistics. Computerized counting can provide early results within minutes of the polls closing, slashing the interval of uncertainty that often spawns rumors or litigation.

This combination of convenience and speed makes e-voting look like the next logical step of the democratic process. But usability alone is not enough. The track record of early online voting pilots is that user-friendly interfaces can cover up basic security flaws. If the voter suspects that malware, insider manipulation, or network failures could taint his or her ballot, the usefulness of convenience quickly disappears. Trust, once lost, is difficult to rebuild. A single high-profile failure or a mystifying technical glitch can overwhelm the news, drown out successful aspects of the system, and keep prospective voters home next election time. In tight elections, even a perception of weakness can invite recounts, legal disputes, and weeks of uncertainty, eroding confidence in the result. In short, emphasizing that convenience is valuable, but without faith in security, it has no value.

Cost volatility adds a second layer of risk. Some public blockchains require a small fee for every transaction, but that fee can rise without warning when network traffic surges. An organization that budgets pennies per vote may discover on election day that the cost has multiplied, forcing unpleasant trade-offs such as rationing transactions, postponing the vote, or absorbing an unexpected financial hit. Smaller jurisdictions, nonprofit associations, and student governments often operate on tight budgets; unpredictability alone can be enough to keep them from considering online options at all. These financial uncertainties make online voting less attractive for smaller organizations, additionally illustrating the importance of systems that balance transparency with predictable, low-cost operations. A reliable system must not only be secure but also economically sustainable in the long term.

Privacy concerns extend beyond the secrecy of the ballot box itself. Every interaction with a website leaves crumbs of data on connection times, kinds of devices, and network addresses that can be reassembled by a devoted analyst. Even if the ballot is encrypted, the surrounding metadata could inform us who voted, when they voted, and even allow us to infer what blocs of voters voted. Through various elections, patterns emerge that unintentionally disclose trends within specific communities or groups. If citizens believe one day their vote can be traced back to them, the chilling effect is immediate. Some do not vote; others self-censor, voting for what appears to be the safest option rather than what they truly prefer. Any viable online system must therefore hide not only the content of each vote but also the behavioral metadata that could reveal how, when, or even whether someone participated. Protecting this invisible layer of information is just as important as protecting the ballot itself.

The broader health of democratic and organizational decision-making depends on solving these intertwined challenges. Whether the setting is a municipal referendum, a corporate merger vote, or a university funding allocation, participants must feel that the act of voting is both accessible and worthwhile. A generalized blueprint that delivers convenience, protects privacy, controls costs, and invites transparent auditing can raise participation rates and strengthen legitimacy across diverse contexts. By confronting these issues head-on, we aim to move online voting from an experimental novelty to a trusted everyday tool. Without addressing them, online voting risks becoming another unfulfilled promise. Which appears attractive on the surface but is unable to gain the trust needed for widespread use.

### *Issues that voters face now*

- Long wait times and travel requirements discourage voter participation.
    - Even motivated voters may give up when they face hours of travel and standing in line, making the process feel more like an obstacle than a civic duty.

- Mail-in ballots can be delayed, lost, stolen, and rejected.
    - Small mistakes like signature mismatches, or late arrivals mean that valid notes are often discarded without the voter's knowledge.

- In-person voting is limited in time and place, halting access.
    - People who have work, family, or health responsibilities can lose interest in voting if they are unable to make it within a limited time frame.

- System errors and system crashes can create suspicion about election results.
    - Even minor glitches can cause doubt, and once the process loses trust, it can be difficult to regain it.
- The idea for our system is to mix convenience with security by using a setup that has a few different parts working together.
    - This combination aims to give voters confidence while keeping the process simple and reliable.

*Actors:*

- **Voters:** People casting their ballots from a computer or phone, who are at the center of the system, and the design ensures their participation is easy, private, and verifiable.

- *Registrar:* The organization responsible for verifying your eligibility to vote and issuing a special one-time ticket. It functions as a gatekeeper, keeping voting separate from identification check-in and not attempting to follow back ballots.

- **Validators:** A small set of machines run by the election board and neutral observers. They store encrypted ballots within the private blockchain while preventing double voting with no single authority to bias the result.

- **Public Blockchain:** An inexpensive public chain that only stores short "fingerprints" (hashes) of votes, with a permanent and tamper-evident record but without divulging sensitive information or being too expensive.

- **Observers/Auditors:** Unbiased third parties who can guarantee the integrity of the process. Their oversight ensures transparency and helps maintain public trust in the system.

*How it works:*

1. **Check-in:** Voter checks in, the registrar confirms they're eligible, and they give them a blind-signed token (like a digital ticket).
    a. This step is significant because this is where the voter and ballot are separated, remaining confidential prior to any vote having even been cast.

2. **Casting a vote:** The voter encrypts their vote, attaches the token, and sends it. Validators make sure the token is not a previously used one, then append the encrypted vote to the private blockchain.

3. **Anchoring:** Every few minutes, validators gather all the new votes, bundle them up, and append one tiny hash of the batch to the public blockchain.
    a. This creates an unalterable digital record that makes tampering impossible to hide.

4. **Verification:** The voter is given a receipt of the public transaction ID and hash of the vote.
    a. It makes it possible for the voters to know for sure that their vote was counted, and it also gives the public a choice to audit the election.

5. **Final tally:** Trustees collectively apply unique keys to decrypt votes, and it can be compared against public hashes. Since decryption is done in conjunction with the involvement of multiple parties, one party cannot corrupt results unilaterally.

This setup tries to hit a few key goals:

- **Anonymity:** Tokens keep votes from being tied back to the voter's name, ensuring that no ballot can be traced back to a specific individual.
    - This protects voters from pressure, retaliation, or unwanted attention that could grow if their choices were revealed. By keeping personal identity separate from the ballot itself, the system encourages free and honest participation.

- **Integrity:** Validators agree on the data, so no single person can cheat or alter results behind the scenes. Each vote is checked by many nodes, and any attempt to tamper with it would be detected immediately.
    - This mutual checking of the votes ensures that the election results reflect the actual votes cast.

- Verifiability: By examining the public anchors published to the blockchain, one can easily verify if tampered votes have occurred. These are permanent fingerprints that show whether a ballot was tampered with.
    - This transparency allows both individual voters and independent auditors to confirm the system's honesty.

- **Lower costs:** Since only the batch fingerprints go public, fees don't get out of control and remain predictable for organizers. By keeping costs low, smaller groups and organizations can reasonably deploy the technology.
    - This approach reduces money costs compared to systems that post each transaction directly to a public blockchain.

## III. RELATED WORK

Many individuals have attempted to implement online voting, and each project, regardless of size, contributes to the overall picture. One of the first academic systems, Helios, proved that a voter could obtain a digital "receipt" and keep the vote secret [1]. It works nicely for student elections and professional societies, but because Helios depends on a single website to publish ballots, a determined attacker or even a simple outage could knock the whole thing offline. That weakness becomes a real concern once turnout grows beyond a few hundred voters. The lesson here is that centralized control, even in a system designed for transparency, creates single points of failure that limit scalability.

Other countries have tested programs like Estonia's national i-Voting program, which demonstrates that citizens will use an online option if it is convenient and integrated into their daily lives [2]. At the same time, security reviews warn that malware on a voter's home computer can quietly change a choice before it reaches the server. This case highlights the inadequacy of strong server-side protections if the voter's own device lacks trustworthiness. The Estonian case highlights how end-to-end security must cover both the infrastructure and the user's environment to be truly reliable.

Switzerland trial-ran sVote, another web-based system, but researchers found ways to trick the tally under certain conditions, leading to a pause in rollout. This case teaches that even well-resourced national systems can falter under deep observation. When weaknesses are discovered after deployment, public trust can quickly vanish, making it harder to try again later.

Mobile-first pilots underscore a clear trade-off. The Voatz app let a small group of overseas and disabled voters submit ballots from their phones. Users loved the convenience, but later studies showed that if a phone was infected or if the data route was tampered with, votes could be altered or revealed [3]. West Virginia attempted to mitigate this risk is by adding a limited blockchain wrapper around Voatz to provide after-the-fact tamper evidence. While the trial ended without disaster, skepticism remained due to lingering doubts about mobile device security and closed-source software. The main takeaway is that usability alone cannot make a system trustworthy if the underlying technology lacks transparency and resilience.

Outside of government, companies and nonprofits have experimented with blockchain to settle board decisions and membership polls. They often appreciated the built-in audit trail, since it reduced disputes over results. However, many quickly discovered the cost curve: posting every single ballot to a busy public chain became financially unsustainable, and the public record risked exposing timing or behavioral data about voters. Some projects tried to mask this metadata with extra cryptography, but the complexity scared away smaller groups without dedicated technical teams. These experiments demonstrate that cost and usability both play the same role as security when it comes to real-world adoption.

Recent research suggests a compromise that mixes the strengths of both worlds. Teams exploring "permissioned" ledgers keep the actual ballots on a short list of trusted computers, then post only a tiny fingerprint to the public chain. Microsoft's open-source Election Guard toolkit, for example, focuses on end-to-end checks that let anyone verify the final count without exposing individual choices [4]. Academic prototypes like E-Vote mix similar ideas with zero-knowledge proofs to cut down on public-chain data. These hybrid designs aim for three simple goals: low cost, clear audit logs, and voter privacy.

Broader studies back up this middle-road thinking. The OECD's 2022 report on blockchain in public services notes that "small, verifiable data anchors" offer most of the security benefit without the storage burden of full on-chain records [5]. Likewise, an MIT Election Lab brief argues that systems that separate "identity management" from "ballot recording" reduce both privacy risk and legal complexity. These insights emphasize the importance of modularity: a system that divides responsibilities between different components is both more secure and easier to manage.

Taken together, these lessons point toward a simple guideline: keep the parts voters can see easy and the parts auditors need solid. The concept in this paper follows that path. It borrows Helios-style verifiability so voters can double-check inclusion, keeps the "can't-fake-it" stamp of a public blockchain, but avoids high fees and data leaks by posting only minimal information. The goal is a realistic, budget-friendly bridge between the ideal of paper-ballot security and the convenience of a phone-based vote.

Reading these case studies shows a pattern: using blockchain only to lock in the final tally solves tampering without driving up costs or exposing ballots. Projects that put every vote on a big public chain got hit by high fees and privacy worries. Simpler hybrid setups worked better. These findings suggest that the balance between transparency and efficiency proved to be more effective, and efficiency is often more important than maximizing decentralization for its own sake. From that we think blockchain is a handy tamper seal, if it is backed by clear rules and technology that anyone can use and is safe.

After considering these examples, it is clear that the real conflict in voting is between decentralization and practicality. Full decentralization achieves full transparency by tracing each and every ballot globally, but at an enormous cost in scalability, user privacy, and cost. Conversely, fully private systems conserve cost and efficiency but risk placing undue trust in a few administrators. The case studies suggest that the hybrid models are not an afterthought but a thoughtful middle ground that is sensitive to the realities of voting: the need for verifiable results without inundating voters or organizers with unnecessary complexity. This compromise is especially worth it when the intended consumers are local communities or small organizations that cannot afford unpredictable costs or specialized infrastructure. While these studies remind us that context is just as crucial as technical design, a system capable of handling millions of votes in a national election may be too large for a university council or nonprofit board, given their limited resources and different stakes.

Conversely, an unadorned private record that might be appropriate to a student government may collapse under the pressure of a high-stakes national election. The projects show that there is no one-size-fits-all model. More feasible is the creation of flexible frameworks that can scale up or down depending on the election size, the data sensitivity level, and resource availability. This context's framing of blockchain allows for solutions applicable to both small organizations and large-scale democratic systems.

## IV. MAIN CONTRIBUTIONS

The most important contribution of this project, in our opinion, is that we went beyond theory and built a working prototype of a blockchain-based voting system. A lot of student projects in this area stop at writing research summaries or drawing diagrams, but we took the next step of writing code, deploying it to a live server, and letting real users interact with it. Our demo, hosted at https://voting-blockchain-50b62.web.app, gives people the chance to see the process in action: logging in, casting a ballot, and verifying that it was counted. Having a real link is important because it proves that online voting isn't just a hypothetical idea; it can run on actual infrastructure with common tools. Beyond that, providing a hands-on example reduces the skepticism that often surrounds purely theoretical papers. It shifts the discussion from "what if" to "here is a running model," to "here is a running model," which is a stronger basis for both critique and improvement.

A second major contribution is that we designed and coded the full end-to-end workflow of an election system. That means we didn't only focus on one part, like the user interface or just the blockchain backend. Instead, we made sure to cover the entire path: a voter checks in, gets a blind-signed token, encrypts their vote, submits it, sees it logged by validators, and then gets a receipt with a unique hash. At the end, all votes can be decrypted together and checked against public records. The end-to-end approach is crucial in voting, as the system collapses if any component is absent. An end-to-end approach is critical in voting because if any piece is left out, the whole system breaks down. Our prototype shows that you can actually connect all the steps into one working flow. By modeling the complete lifecycle, we also created a foundation for stress testing and scaling, since future researchers could analyze which stages are most likely to face bottlenecks or attacks.

Another key contribution is in the technological choices we made. We didn't rely on obscure or academic-only software. Instead, we built the frontend using Next.js and React, with TailwindCSS for styling and TypeScript for type safety and cleaner code. The backend used Firebase for authentication, database storage, and hosting. This stack is accessible and familiar to a lot of student developers, which means others could actually reproduce our results and keep experimenting. It also demonstrates that secure systems don't always need complicated or expensive enterprise infrastructure. By using mainstream web technologies, we made blockchain voting more approachable for students, small organizations, and even local communities that might want to try it. This choice was intentional: lowering the technical barrier is just as important as securing the system, because a tool that nobody can adopt fails its practical mission.

On the blockchain side, our main contribution was implementing a hybrid model. Many designs try to put every single vote on a public blockchain. That sounds nice for transparency, but it's totally impractical when you think about costs and scalability. We solved this by building a two-layer system: a private blockchain for the encrypted ballots and then periodic anchoring to a public blockchain by storing only a hash of the recent batch. This balances cost, speed, and trust. The private chain keeps the detailed records safe and manageable, while the public chain acts like a permanent seal that prevents tampering. This represents a significant contribution because it demonstrates that you don't have to choose between options that are either "secure but expensive" or "cheap but weak." With careful design, you can get both. The hybrid approach also opens opportunities for modular upgrades in the future; for example, organizations could swap in a different public chain for anchoring without changing the rest of the system.

We also contributed to the user experience side of online voting. A lot of secure systems end up scaring users away because the steps are too technical. We wanted to prove that a blockchain-based system could still be simple and accessible. Our site works in any browser with no extra software required. The interface is straightforward, mobile-friendly, and built with accessibility in mind, so it supports screen readers, large fonts, and flexible layouts. For us, this contribution was about proving that "secure" doesn't have to mean "unusable." In fact, we think that ease of use is just as important as cryptography when it comes to getting people to trust and use a system. As well, in practice, a system that feels natural to voters is more likely to increase attendance and reduce errors while still maintaining strong back-end security.

An additional contribution was how we designed the web interface as a live teaching and testing tool. The top of the page gives voters a simple input box and a "Vote" button, so anyone can try the system right away. Underneath, the interface splits into two sections: Current Votes, which shows each candidate's name and the current vote totals in real time, and Blockchain Data, which lists every block in the chain with its index, timestamp, candidate name, hash, and previous hash. This dual view makes the system unique: a regular voter can just check the totals, while a more technical user can scroll down and verify the cryptographic chain of blocks. This contribution is important because it shows that blockchain transparency can be presented in a way that's understandable to both everyday users and more advanced observers. By designing transparency into the interface itself, we reduce reliance on outside experts to explain how the system works.

Another big contribution was testing the challenges of blockchain costs and performance in practice. Instead of writing "transaction fees could be a problem" and leaving it at that, we had to deal with them when coding. Public blockchain fees can spike randomly, and if every single vote were a transaction, the system would collapse financially. Our batching method was a real solution to this issue, and testing it in code gave us data to show that it works. This kind of hands-on problem-solving is a contribution because it grounds research in real-world limitations rather than just ideal models. More importantly, it allowed us to highlight how cost structures directly affect adoption, something that often is overlooked in technical proposals.

We also see our project as a contribution to the educational side of online voting research. Because our code is live, open, and hosted online, it can be used by other students or researchers as a starting point. Someone could fork our project to test new cryptographic methods, swap it in a different blockchain, or even scale it up for a classroom election. Having a live demo plus working code gives people something concrete to build on, instead of starting from scratch. By leaving behind reusable tools, our project continues contributing even after the paper is finished.

Finally, another contribution is how we framed the system for different contexts. While national elections are the long-term goal, we intentionally designed our project so it could be theoretically useful right now for smaller groups' purposes, like universities, nonprofit boards, student governments, or company shareholder votes. These organizations have many of the same problems (low turnout, logistics, costs), but they don't have the same budget or infrastructure as a government agency. By showing that our system works in these smaller contexts, we contribute a more realistic and immediate path for blockchain voting to be adopted. Our design philosophy is that innovation should first serve those who can realistically implement it, instead of only aiming for massive national deployments.

In summary, our main contributions include:

- Building a live, working prototype that anyone can test.
    - This makes the project tangible and shows that the concept works outside of theory.

- Implementing a full end-to-end voting workflow in code.
    - This ensures every stage of the voting process is covered, so nothing critical is left out.

- Using accessible, common web tools (Next.js, Firebase, TypeScript) to make the system reproducible.
    - This lowers the barrier for future students and organizations to replicate or extend the project.

- Designing a hybrid blockchain model that balances cost and security.
    - The two-layer system shows that adaptability and trust can coexist.

- Prioritizing usability and accessibility alongside cryptographic security.
    - A user-friendly interface ensures that the system is practical, not just secure.

- Creating a real web interface that lets people cast a test vote and then view the blockchain data directly.
    - This dual view serves both casual voters and technical observers.

- Testing blockchain costs and performance through real implementation.
    - Hands-on testing gave concrete data about fees and potential system limitations.

- Providing an educational platform for future students and researchers.
    - The open codebase and live demo create opportunities for continued learning and experimentation.

- Showing how blockchain voting can realistically be applied to small-scale organizations today.
    - This provides a clear path for immediate, practical adoption before scaling to larger elections.

**Blockchain Voting System Prototype**

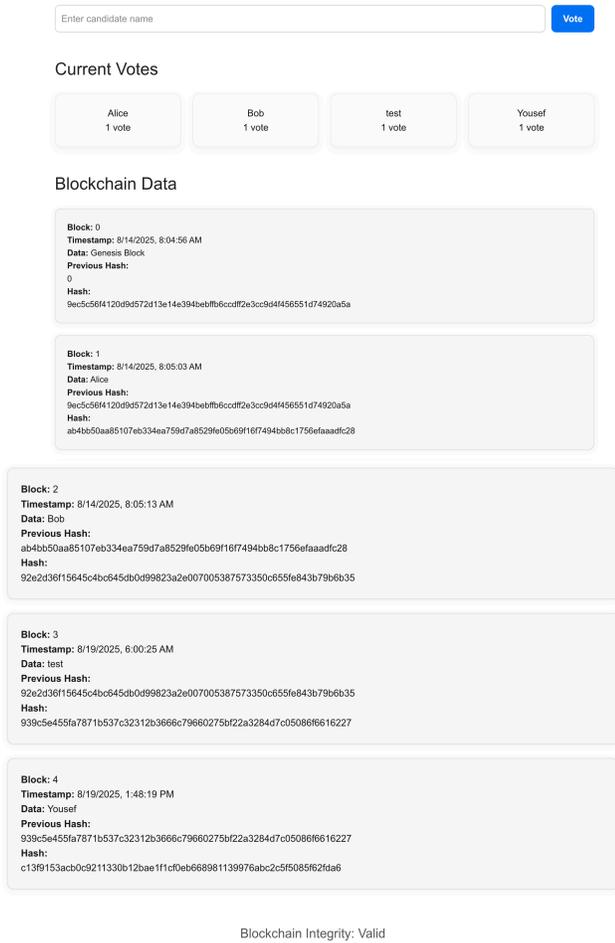

Figure 1: Blockchain voting system interface prototype demonstrating backend data.

## V. SUMMARY & CONCLUSION

Online voting has always promised shorter lines and better access, but many attempts have struggled to earn public trust. Our concept builds on what past research has shown works well while avoiding some of the more expensive or risky choices. By blending familiar online tools with a minimal use of blockchain, we can protect privacy, ensure votes are counted accurately, and keep costs manageable. This balance is not just a technical success but a policy-oriented one: practical adoption depends on the ability to address concerns from both engineers and the public at once.

This kind of system could help a wide range of groups, like local governments, student councils, or small companies, run trustworthy elections without needing a lot of special equipment or training. The main data remains private while the important proof is shared publicly, making it easier for outside observers to confirm results without seeing the actual votes. That dual visibility—private ballots but public proofs—creates a stronger baseline of trust than either extreme model. In contexts where disputes often arise over legitimacy, that kind of layered verification is essential.

We also looked at several existing projects in this space. From Helios to mobile-based pilots like Voatz, and even studies from organizations like the OECD and MIT Election Lab, there is a growing interest in combining voter privacy with clear verification. Many of these earlier projects helped shape our thinking by showing what works, what doesn't, and where the gaps still remain. Our idea tries to learn from what succeeded in these experiments and what they failed to achieve, both technically and in terms of public confidence. By placing our work in that tradition, we are recognizing that innovation in online voting is incremental: each model draws on the experiences of the previous one.

The possible difference between those past projects and ours is that we didn't stop at the theoretical stage. Instead of just describing what such a system could be, we proceeded to actually make a working prototype. We wrote the code, deployed it online, and made it accessible at a live project link where anyone can interact with it, https://voting-blockchain-50b62.web.app. This contribution moves the discussion from "concept" to "practice." People can now open a webpage, cast a ballot, and see the result anchored through blockchain verification. That alone shows that a system like this can exist without requiring a massive budget or national-level resources. More importantly, it provides an interactive foundation for critics and supporters alike to test, evaluate, and debate.

Another major contribution was demonstrating a full end-to-end workflow. Our prototype doesn't just collect inputs; it handles the whole lifecycle of a vote: the voter check-in, token assignment, encryption, submission, validation, receipt generation, and verification through hashes on a blockchain. By connecting these steps together in code, we provided a practical example of how the pieces fit as a complete system. This is valuable because most past work tends to isolate pieces (like cryptographic proofs or ballot interfaces) instead of tying them together. Having a unified workflow makes it easier to see how weaknesses in one stage might ripple outward, creating vulnerabilities elsewhere.

The interface design also became an important part of our contribution. We realized early that if an online voting platform feels confusing, no amount of security will matter. So, we made the webpage straightforward and easy to use. A voter can log in, select a candidate, submit their vote, and instantly receive a digital confirmation. More importantly, the interface doesn't just say "thank you"—it provides receipts and lets voters see how their vote is represented in the system. For testing, anyone can perform a demo vote and actually view how the data appears, giving a clear picture of transparency. This hands-on feature turns the idea into something users can trust. The takeaway is that usability is a form of security and an overlooked but critical truth in system design.

We also showed that you do not need unusual or obscure technology to make something like this work. Our system was built using Next.js, React, TypeScript, TailwindCSS, and Firebase, which are all common in today's web development. This feature makes the project replicable for other students or organizations. Someone with intermediate coding knowledge can follow our structure, run their own demo, or even adapt it for local use. That practicality is itself a contribution, as far too many blockchain projects are still out of reach for regular developers. By making it easier to enter, we stand a better chance that future researchers or civic groups can take our model and advance it instead of having to start from scratch.

On the blockchain side, we didn't go to extremes. Instead of putting every single vote directly onto a public chain—which would be slow and expensive—or hiding everything in a private database, we used a hybrid approach. Votes are stored in a private chain for efficiency, while public hashes are recorded on an open blockchain to guarantee tamper evidence. This design balances cost, transparency, and speed. Our testing confirmed that it is a realistic compromise, especially for smaller elections that don't have the resources of a national government. By demonstrating a middle path, we also encourage more nuanced discussions: debates about blockchain voting are too often framed as "all or nothing," when in fact layered systems may be the most workable solution.

Perhaps one of the most lasting contributions is that we left behind something concrete: a working demo and real code. Future students or researchers can copy the project, experiment with it, and expand on it. A college student council could use it to plan campus elections, or a nonprofit organization could use it for elections to boards. By leaving behind something of a living, working example, we created more than an article; we created something others could use as a starting point to expand on.

In short, our research contributes several things to the ongoing debate on secure online voting. We showed a live and functional prototype, we turned theory into practice by implementing the whole voting process workflow, we designed an interface that proves that transparency and usability are not rivals, we tested a hybrid blockchain solution that balances practicality and trust, and we made the project reproducible to others with a familiar tech stack. These steps illustrate that blockchain-based voting is not a far-fetched concept; it is possible to experiment and test today on working, real systems. Though there is still a lot to be done, our project provides evidence that students can effectively contribute to this area through coding, deploying, and making available systems that can be experimented with by others. More broadly, it confirms the hypothesis that progress toward secure online voting will not be achieved through a single breakthrough but gradually through many prototypes, each approximating a slightly different facet of the problem.